\documentclass[12pt,onecolumn]{article}
\usepackage{setspace}
\usepackage[utf8]{inputenc}
\usepackage{amsmath,amsthm,amssymb,amsfonts,mathtools,MnSymbol}
\usepackage{amssymb}
\usepackage[inline]{enumitem}
\usepackage{geometry}
\usepackage{caption}
\usepackage{subcaption}
\usepackage{natbib}
\usepackage{authblk}
\usepackage{graphicx}
\usepackage{color}
\usepackage{diagbox}
\def\e{\mathrm{e}}

\providecommand{\keywords}[1]
{
  \small	
  \textbf{\textit{Keywords---}} #1
}

\title{An Euler-Bernoulli-Type  Beam Model of the Vocal Folds for Describing Curved and Incomplete  Glottal Closure Patterns}

\author[1]{Mohamed A. Serry}
\affil[1]{Mechanical and Mechatronics Engineering, University of Waterloo, Waterloo, Ontario N2L 3G1, Canada}
\author[2]{Gabriel A. Alzamendi}
\affil[2]{Institute for Research and Development on Bioengineering and Bioinformatics (IBB), CONICET-UNER, Oro Verde,
Entre R\'ios 3100, Argentina}
\author[3]{ Mat\'ias Za\~nartu}
\affil[3]{Department of Electronic Engineering, Universidad T\'ecnica Federico Santa Mar\'ia, Valpara\'iso, Chile}
\author[1]{Sean D. Peterson}%\email{peterson@uwaterloo.ca}
\date{}
\begin{document}
%\twocolumn[
 % \begin{@twocolumnfalse}
\maketitle
\begin{abstract}
Incomplete glottal closure is a laryngeal configuration wherein the glottis is not fully obstructed prior to phonation. It has been linked  to inefficient voice production and voice disorders. Various incomplete glottal closure patterns can arise and the mechanisms driving them are not well understood. In this work, we introduce an Euler-Bernoulli composite beam vocal fold (VF) model that produces qualitatively similar incomplete glottal closure patterns as those observed in experimental and high-fidelity numerical studies, thus offering insights in to the potential underlying physical mechanisms. Refined physiological insights are pursued by incorporating the beam model into a VF posturing model that embeds the five intrinsic laryngeal muscles. Analysis of the combined model shows that co-activating the lateral cricoarytenoid (LCA) and interarytenoid (IA) muscles without activating the thyroarytenoid (TA) muscle results in a bowed (convex) VF geometry with closure at the posterior margin only; this is primarily attributed to the reactive moments at the anterior VF margin. This bowed pattern can also arise during VF compression (due to extrinsic laryngeal muscle activation for example), wherein the internal moment induced passively by the TA muscle tissue is the predominant mechanism. On the other hand, activating the TA muscle without incorporating other adductory muscles results in anterior and mid-membranous glottal closure, a concave VF geometry, and a posterior glottal opening driven by internal moments induced by TA muscle activation. In the case of initial full glottal closure, the posterior cricoarytenoid (PCA) muscle activation cancels the adductory effects of the LCA and IA muscles, resulting in a concave VF geometry and posterior glottal opening. Furthermore, certain maneuvers involving co-activation of all adductory muscles result in an hourglass glottal shape due to a reactive moment at the anterior VF margin and moderate internal moment induced by TA muscle activation. These findings have implications regarding potential laryngeal maneuvers in patients with voice disorders involving imbalances or excessive tension in the laryngeal muscles such as muscle tension dysphonia.
\end{abstract}

\keywords{Vocal folds; Incomplete glottal closure; Muscle tension dysphonia; Euler-Bernoulli beam.}
%  \end{@twocolumnfalse}
%]

%%%%%%%%%%%%%%%%%%%%%%%%%%%%%%%%%%%%%%%%%%%%%%%%%%%%%%%%%%%%%%%%%%
\section{Introduction}
The configuration of the vocal folds (VFs), a cornerstone of voice production, is determined by the particular combination of activated intrinsic and extrinsic laryngeal muscles. Nominally, the VFs are completely adducted prior to the onset of phonation, and their interaction with the air flow driven by the lungs results in vibrations and consequent acoustic waves, which forms the basis of voiced speech. In  some  scenarios, complete glottal closure is not attained, which can result in inefficient voice production \citep{ZanartuGalindoErathPetersonWodickaHillman14}, and, in some cases, stress concentrations in the VFs that may lead to VF trauma \citep{DejonckereKob09}. Hence, incomplete glottal closure is often linked to disorders that are associated with inefficiencies in, or damage to, the vocal mechanism, including Parkinson's disease \citep{HansonGerrattWard84}\footnote{Parkinson’s disease is a relatively  prevalent  disorder, affecting the human central, peripheral, and enteric nervous systems \citep{BraakBraak00}.} and muscle tension dysphonia (MTD) \citep{MorrisonRammage93}\footnote{MTD \citep{MorrisonRammage93}, also known as non-phonotraumatic hyperfunction \citep{HillmanSteppVanStanZanartuMehta20}, is a class of voice disorders associated with misuse of the vocal mechanisms without the presence of organic changes in the vocal organs, leading to low speech quality and vocal fatigue, with  a wide range of symptoms and patterns, including excessive/unbalanced activation of intrinsic and extrinsic laryngeal muscles \citep{Roy08,HocevarBoltezarJankoZargi98}, supraglottal compression \citep{MorrisonRammage93}, and abnormal fundamental frequency  \citep{NguyenKennyTranLivesey09,AltmanAtkinsonLazarus05}.}.

\begin{figure}[htpb]
     \centering
     \includegraphics[width=0.5\linewidth]{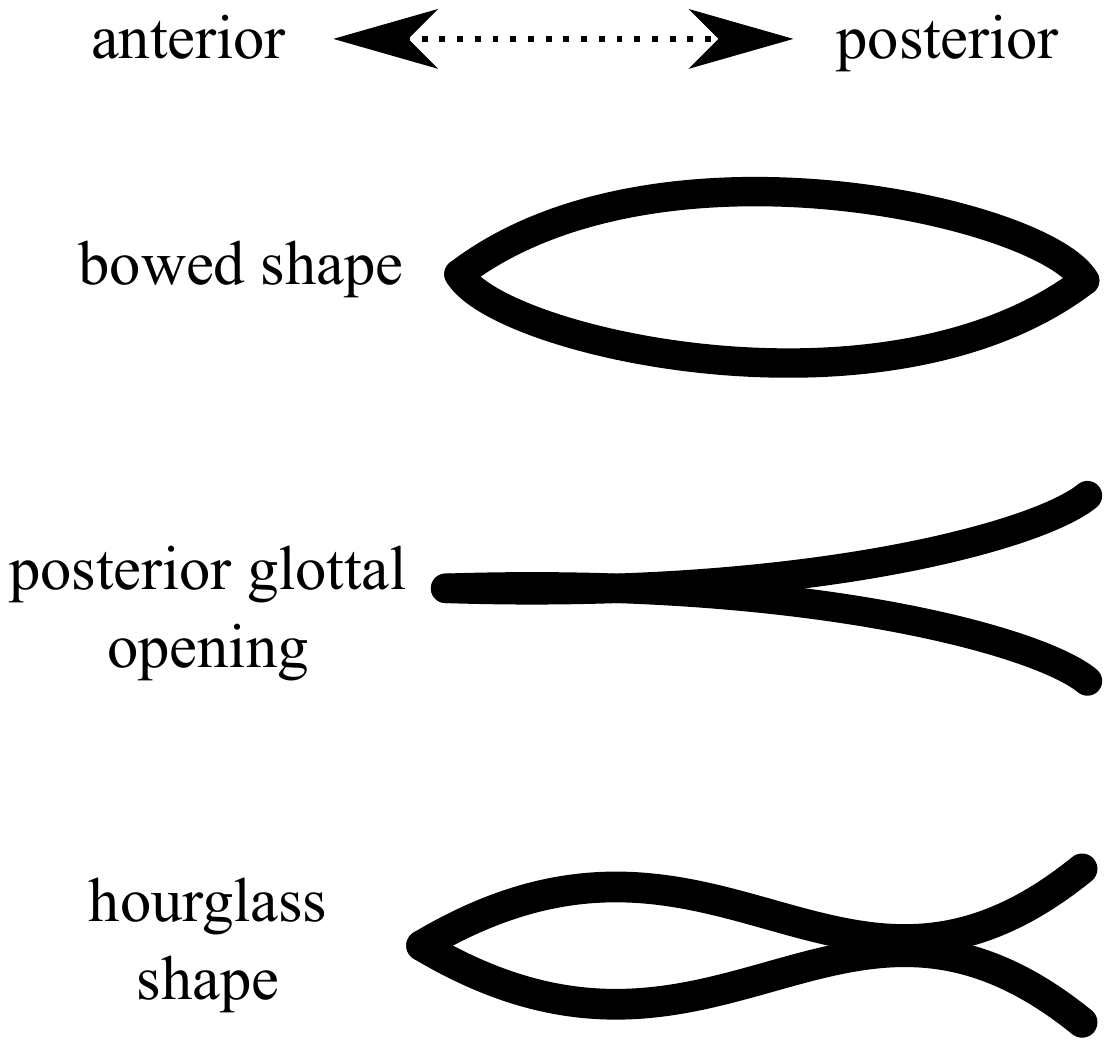}
     \caption{Schematic diagram (superior view) of some incomplete and curved glottal closure patterns that are observed clinically.}
     \label{fig:GlottalShapes}
 \end{figure}

 Incomplete glottal closure\footnote{In this study we refer to the glottal configuration of the VFs at rest immediately prior to phonation initiation, identifying any gaps between the folds as incomplete glottal closure. In clinical settings, incomplete glottal closure typically refers to gaps between the folds when the VFs are at their maximum glottal closure phase during phonation \citep{SoderstenHertegardHammarberg95,NguyenKennyTranLivesey09}. Our definition herein isolates laryngeal factors, which are the focus of this study, by dismissing the dynamics of VF vibrations.} comes in various patterns   \citep{MorrisonRammage93,NguyenKennyTranLivesey09,SoderstenHertegardHammarberg95} as shown schematically in Figure \ref{fig:GlottalShapes}, including bowed shape: a glottal pattern wherein the left and right VF geometries are convex with a gap at the mid-membranous portion; posterior glottal opening: full glottal closure is achieved in the anterior and mid-membranous regions only, leaving the posterior margin open; and hourglass glottal configuration: a pattern  with  anterior and posterior gaps and potential VF contact in the mid-membranous region. There exist other incomplete glottal closure patterns, sharing similarities with those mentioned above, such as spindle-shaped glottis, and anterior opening (see \citet{RajaeiBarzegarMojiriNilforoush14,SoderstenHertegardHammarberg95}). These latter patterns are not addressed in this work.
 
%  Therefore, it is important to understand the mechanisms, underlying these patterns of  incomplete glottal closure.

There exist several experimental and clinical studies in the literature that attempt to elucidate, at least in part, some of the laryngeal mechanisms associated with curved and incomplete glottal closure patterns. Based upon inspection of cadaver larynges, \citet{MorrisonRammage93} posited that a posterior glottal opening is associated with excessive activation of the posterior cricoarytenoid (PCA) muscle. \citet{ChoiYeBerkeKreiman93} conducted experimental investigations of excised canine models and found that activating the thyroarytenoid (TA) muscle, while keeping other adductory laryngeal muscles inactive, leads to anterior and mid-membranous glottal closure, whereas the posterior glottis stays open, thus resulting in a closure pattern similar to a posterior glottal opening. On the other hand, they found that when the TA muscle is relaxed and other adductory muscles (LCA/IA) are activated, closure is achieved only at the posterior margins of the VFs with mid-membranous opening \citep{ChoiYeBerkeKreiman93,ChhetriNeubauer15}, thus leading to a bowed configuration as seen in Figure \ref{fig:GlottalShapes}. Moreover, complete glottal closure of excised canine larynges was attained via co-activation of all adductory muscles. More recent clinical investigations using refined experimental setups  (see, e.g., \citet{ChhetriNeubauer15}) further confirm these observations. \citet{ChhetriNeubauerBerry12} conducted a parametric study of the effects of intrinsic laryngeal muscle activation, modulated by graded stimulation, on the pre-phonatory posture of a canine model. In addition to confirming the findings of \citet{ChoiBerkeYeKreiman93}, \citet{ChhetriNeubauerBerry12} found that when keeping the TA muscle activation at a constant level and increasing activation of the cricothyroid muscle (CT), the glottal area increases and the medial bulging caused by the TA muscle activation is reduced. In addition, they observed that when keeping the LCA and IA muscle activation at constant levels and increasing CT muscle activation, the glottis starts to open posteriorly and the glottal area increases. Interestingly, the authors found that with certain muscular executions involving co-activation of the LCA, IA, and TA muscles, the  glottis exhibits an hourglass shape  (see \citet[Fig.~10]{ChhetriNeubauerBerry12}).  

Besides the aforementioned clinical and experimental works, there exist numerical studies that shed some light onto curved and incomplete glottal closure patterns. \citet{HunterTitzeAlipour04} developed one of the early three-dimensional VF posturing models, where adductory and abductory muscles are incorporated, showing that full activation of the LCA muscle induces nonuniform curvature of the medial surface. \citet{DejonckereKob09} studied the influence of incomplete glottal closure patterns (with linear and curved VF geometries) on VF vibrations using a multi-mass model, showing that some resting incomplete glottal closure configurations may induce localized VF impact, which they hypothesized to be a potential underlying mechanism inducing VF trauma, especially in females. However, the authors did not study the laryngeal maneuvers that induce these resting glottal shapes. \citet{YinZhang14,YinZhang16} conducted numerical simulations using high-fidelity numerical models, showing that posterior glottal opening occurs with the sole activation of the TA muscle, mid-membranous opening when the LCA and IA muscles are co-activated (without incorporating the TA muscle), and full glottal closure when all adductors are co-activated, in agreement with the aforementioned clinical observations. In a more recent study, \citet{GengPhamXueZheng20} proposed a detailed physiologically accurate finite-element posturing model,   based on MRI scan images of a canine larynx. Even though the study does not study the glottal geometry, it provides useful insights into  how synergistic activation of laryngeal muscles exhibits complex interaction with laryngeal variables (e.g., VF strain, rotation and translation of arytenoid cartilages, and glottal area).    Recently, research interest has also been directed towards investigating how activating laryngeal muscles alters the VF medial surfaces \citep{PillutlaReddySchlegelZhangChhetri22}.

Despite these valuable efforts, a clear picture of the physical mechanisms inducing  different glottal patterns remains elusive. It is challenging to isolate and control the factors underlying posturing mechanics experimentally. Moreover, high-fidelity numerical models, despite their accuracy in replicating physiological laryngeal postures, do not provide clear intuitive understanding of the mechanics of posturing, and typically suffer from high computational costs. We hypothesize that the non-homogeneous structure of the VFs, which comprise overlapping tissue layers with different mechanical and geometrical properties \citep{TitzeAlipour06}, underlies, in part, the different glottal shapes displayed in Figure \ref{fig:GlottalShapes}. As such, we propose an Euler-Bernoulli composite beam model of the VFs to elucidate some of the mechanisms underlying glottal patterns prior to phonation. Beam models have been utilized previously to explore VF vibrations and phonation fundamental frequency \citep{TitzeHunter04,ZhangSiegmundChan07}. We opt for this relatively simple modeling framework to facilitate exploration of the mechanisms underlying the resulting  glottal shapes. To gain refined physiological insights into how intrinsic laryngeal muscles may influence glottal geometry, the proposed model is integrated with the muscle-controlled posturing model of \citet{TitzeHunter07}.  
%Our model  is capable of providing  useful insights  into the mechanisms underlying different glottal shapes and the relevance of these mechanisms in  voice disorders, including MTD. %Moreover, our modelling framework, due to its relative simplicity in comparison to high-fidelity models and reasonable accuracy, can be integrated in future works with numerical phonation models to gain better understanding of the mechanics of voiced speech. 

The organization of this work is as follows: a detailed derivation of the composite beam VF model is introduced in Section \ref{sec:ModelDerivation}; analysis is conducted in Section \ref{sec:TheoreticalAnalysis}; numerical simulations of the integrated beam and posturing model are presented in Section \ref{sec:Simulations}; Section \ref{sec:Discussion} presents discussion of the results; and the study is concluded in Section \ref{sec:Conclusion}.

%%%%%%%%%%%%%%%%%%%%%%%%%%%%%%%%%%%%%%%%%%%%%%%%%%%%%%%%%%%%%%
\section{Model development}
\label{sec:ModelDerivation}
\begin{figure}[htpb]
    \centering
    \includegraphics[width=0.5\linewidth]{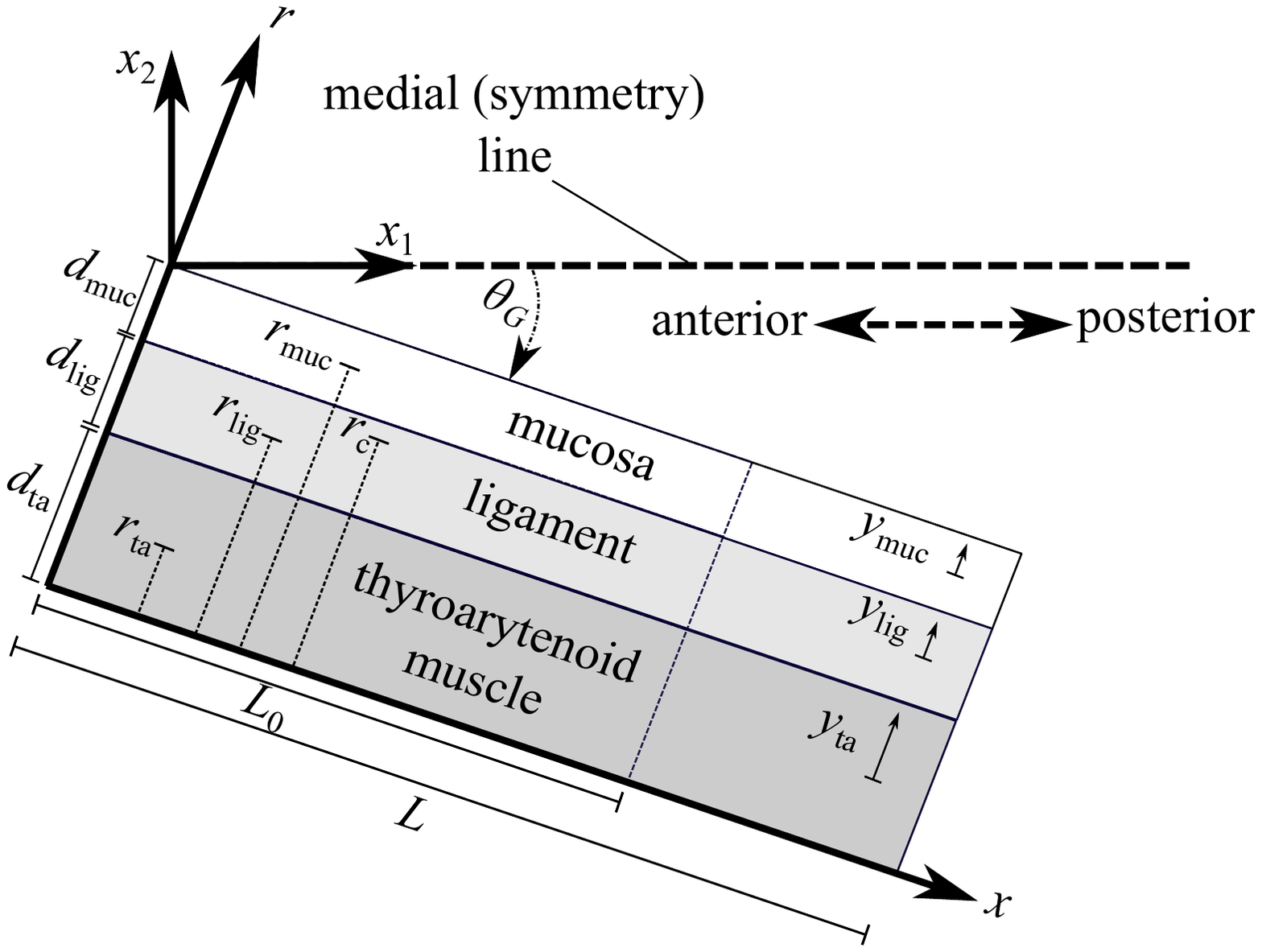}
    \caption{Schematic diagram (superior view) of the VF composite beam model.}
    \label{fig:SchematicBeam}
\end{figure}

Herein, we propose a static Euler-Bernoulli-type composite beam model (see, for example, \citet{BauchauCraig09}) for the VFs, with the different VF layers represented by strata in the beam. For simplicity we assume symmetry with respect to the medial plane; hence, we consider only one (the left) VF. A schematic representation of the composite beam model is shown in Figure \ref{fig:SchematicBeam}. 

The VF beam model consists of three layers: (1) the mucosa with depth $d_\mathrm{muc}$ and cross-sectional area $A_\mathrm{muc}$; (2) the vocal ligament with depth $d_\mathrm{lig}$ and cross-sectional area $A_\mathrm{lig}$; and (3) the thyroarytenoid (vocalis) muscle  with depth $d_\mathrm{ta}$ and cross-sectional area $A_\mathrm{ta}$.  For the sake of compact presentation, we define the index set
\begin{equation}
    \mathcal{I}=\{\mathrm{muc},\mathrm{lig},\mathrm{ta}\},
\end{equation}
where $\mathrm{muc}$, $\mathrm{lig}$, and $\mathrm{ta}$ refer  to the mucosa, ligament, and TA muscle tissue, respectively. We assume that each layer has a  uniform rectangular cross-section and the layer thicknesses (in the inferior-superior direction) are equal and denoted by $b$; thus, layer depth can be computed as $d_i = A_{i}/ b$ for $i \in \mathcal{I}$.  

Our modelling framework assumes that VF deformation consists of (a) potentially large longitudinal stretching/compression with uniform strain, and (b) modest bending  due to the induced moments inside the VF.
% \begin{figure}
%     \centering
%     \includegraphics[width=0.95\linewidth]{Deformation.pdf}
%     \caption{Schematic diagram of the deformation of the VF model.}
%     \label{fig:Deformation}
% \end{figure}
Let $L_{0}$ denote the resting VF length and $L$ denote the VF length after longitudinal deformation due to the associated nominal uniform strain $\bar{\varepsilon}$; that is,
\begin{equation}\label{eq:VFElongation}
L=(1+\bar{\varepsilon})L_{0}.
\end{equation}
We assume the nominal strain is known a priori\footnote{Such as from the two-dimensional posturing model of \citet{TitzeHunter07}, which incorporates the mechanics of the arytenoid cartilages and cricothyroid joints, and relates them to VF strain, see the discussion in Section \ref{sec:Simulations}.}.% Therein,  the VF is assumed to be a one-dimensional spring-like element; hence,  the posturing model in  \citet{TitzeHunter07} does not account for moment equilibrium.

Let $x\in [0,L]$ denote the position along the deformed VF configuration (after applying  strain $\bar{\varepsilon}$) relative to the anterior VF margin, and $r$ denote the depth position along the axis perpendicular to the VF axis relative to the base of the TA muscle (see Figure \ref{fig:SchematicBeam}). Consider a plane VF cross-section at position $x$, and let $y_\mathrm{muc}$ denote the relative  position along the $r$-axis with respect to the geometrical center of the mucosa (i.e., $y_\mathrm{muc}=0$ corresponds to the geometrical center of the mucosal cross-section). Similarly, let $y_\mathrm{lig}$ and $y_\mathrm{ta}$ be analogous coordinates for the ligament and TA muscle, respectively (see Figure \ref{fig:SchematicBeam}). Note that the range of $y_{i}$ is $[-d_{i}/2,d_{i}/2]$ for $i\in \mathcal{I}$.  

Let $w(x)$ denote the transverse deflection of the beam (in the $r$-direction). Moreover, let $u_{i}(x,y_{i})$, $i\in \mathcal{I}$,  denote the longitudinal displacement of the $i^{\mathrm{th}}$ VF layer, where longitudinal displacements are with respect to the  deformed VF configuration under $\bar{\varepsilon}$. In addition, let $\bar{u}_{i}(x)=u_{i}(x,y_{i}=0)$, $i\in \mathcal{I}$, denote the   longitudinal displacement at the center of the $i^{\mathrm{th}}$ layer.  Under Euler-Bernoulli beam theory (see, for example, \citet{BauchauCraig09}), the longitudinal displacement functions can be written as 
\begin{equation}\label{eq:DisplacementField}
u_{i}=\bar{u}_{i}-y_{i}w',~ i\in \mathcal{I},
\end{equation}
% \begin{align}
% u_\mathrm{muc}(x,y_\mathrm{muc})&=\bar{u}_\mathrm{muc}(x)-y_\mathrm{muc}w',~y_\mathrm{muc}\in [-d_\mathrm{muc}/2,d_\mathrm{muc}/2]\\
% u_\mathrm{lig}(x,y_\mathrm{lig})&=\bar{u}_\mathrm{lig}(x)-y_\mathrm{lig}w',~y_\mathrm{lig}\in [-d_\mathrm{lig}/2,d_\mathrm{lig}/2]\\
% u_\mathrm{ta}(x,y_\mathrm{ta})&=\bar{u}_\mathrm{ta}(x)-y_\mathrm{ta}w',~y_\mathrm{ta}\in [-d_\mathrm{ta}/2,d_\mathrm{ta}/2],
% \end{align}
%
where the prime symbol denotes differentiation with respect to $x$.  Continuity of displacement fields necessitates that 
%\begin{align*}
$u_\mathrm{muc}(x,-d_\mathrm{muc}/2)= u_\mathrm{lig}(x,d_\mathrm{lig}/2)$ and $
u_\mathrm{lig}(x,-d_\mathrm{lig}/2)= u_\mathrm{ta}(x,d_\mathrm{ta}/2)$ for all $x\in [0,L]$,
%\end{align*}
which yields the conditions
% \begin{align*}
% \bar{u}_\mathrm{lig}&=\bar{u}_\mathrm{muc}+\frac{1}{2}(d_\mathrm{lig}+d_\mathrm{muc})w',\\
% \bar{u}_\mathrm{ta}&=\bar{u}_\mathrm{lig}+\frac{1}{2}(d_\mathrm{ta}+d_\mathrm{lig})w',
% \end{align*}
% and consequently,
%
\begin{equation}\label{eq:ContinuityCondition}
\begin{split}
\bar{u}_\mathrm{lig}&=\bar{u}_\mathrm{muc}+\frac{1}{2}(d_\mathrm{lig}+d_\mathrm{muc})w',\\
\bar{u}_\mathrm{ta}&=\bar{u}_\mathrm{muc}+\frac{1}{2}(d_\mathrm{ta}+2d_\mathrm{lig}+d_\mathrm{muc})w'.
\end{split}
\end{equation}

Given longitudinal displacement in the $i^\mathrm{th}$ layer with respect to the deformed configuration under $\bar{\varepsilon}$, the total strain in that layer is given by\footnote{Consider an infinitesimal line element $\mathrm{d}x_{0}$ that experiences a composition of two deformations: the first is longitudinal deformation with associated uniform normal  strain $\varepsilon_{0}$, and the second is due to a longitudinal displacement field $u$ (with respect to the configuration after applying the strain $\varepsilon_{0}$).  The length of the line element after applying  strain $\varepsilon_{0}$ is $\mathrm{d}x_{1}=(1+\varepsilon_{0})\mathrm{d}x_{0}$ ($\mathrm{d}x_{0}=\mathrm{d}x_{1}/(1+\varepsilon_{0})$) and the length after applying the displacement field is $\mathrm{d} x_{2}=(1+\mathrm{d}u/\mathrm{d}x_{1})\mathrm{d}x_{1}$. Therefore, the total strain due to the combination of the two deformations is $\varepsilon=(\mathrm{d}x_{2}-\mathrm{d}x_{0})/\mathrm{d}x_{0}=\varepsilon_{0}+(1+\varepsilon_{0})\mathrm{d}u/\mathrm{d}x_{1}$.}
\begin{equation}\label{eq:GeneralStrainEquation}
\varepsilon_{i}=\bar{\varepsilon}+(1+\bar{\varepsilon})u'_{i},~i\in \mathcal{I}.
\end{equation}
Substituting Equation \eqref{eq:DisplacementField} into Equation \eqref{eq:GeneralStrainEquation} results in
\begin{equation}\label{eq:StrainField}
    \varepsilon_{i}=\bar{\varepsilon}+ (1+\bar{\varepsilon})\left(\bar{u}'_{i}-y_{i}w''\right),~i\in \mathcal{I}.%~y_{i}\in \left[-\frac{d_{i}}{2},\frac{d_{i}}{2}\right],
\end{equation}
%
% \begin{align}
% \varepsilon_\mathrm{muc}(x,y_\mathrm{muc})&=\bar{\varepsilon}+ (1+\bar{\varepsilon})\left(\bar{u}'_\mathrm{muc}(x)-y_\mathrm{muc}w''\right),\\
% \varepsilon_\mathrm{lig}(x,y_\mathrm{lig})&=\bar{\varepsilon}+(1+\bar{\varepsilon})\left(\bar{u}'_\mathrm{lig}(x)-y_\mathrm{lig}w''\right),\\
% \varepsilon_\mathrm{ta}(x,y_\mathrm{ta})&=\bar{\varepsilon}+(1+\bar{\varepsilon})\left(\bar{u}_\mathrm{ta}(x)-y_\mathrm{ta}w''\right).
% \end{align}
%$i\in \mathcal{I},~x\in [0,L]$.
The stress field is estimated from strain and, in the case of the TA muscle layer, TA muscle activation $\mathtt{a}_\mathrm{ta}$, which 
is a non-dimensional parameter, ranging between 0 and
1, that corresponds to the activation level in the
TA muscle, with 0 indicating a completely flaccid muscle and 1 being maximum contraction. Herein,  we utilize local linearization about the nominal strain $\bar{\varepsilon}$. That is, the stress functions in the VF layers, $\sigma_{i},~i\in \mathcal{I}$, are given by the approximate relations
% $$
% \sigma_{i}(\varepsilon)\approx \bar{\sigma}_{i}(\bar{\varepsilon})+E_{i}(\varepsilon-\bar{\varepsilon}),
% $$
% where
% $
% E_{i}=\mathrm{d}\bar{\sigma}_{i}/\mathrm{d} \varepsilon|_{\varepsilon=\bar{\varepsilon}}
% $.
% By applying this reasoning to each VF layer, we obtain the approximate stress relations:
%
\begin{equation}\label{eq:StressField}
\sigma_{i}= \sigma_{i,0}+E_{i}(\varepsilon_{i}-\bar{\varepsilon}),~i\in \mathcal{I}, 
% \begin{split}
% \sigma_\mathrm{muc}(\varepsilon_\mathrm{muc})&= \sigma_{\mathrm{muc},0}+E_\mathrm{muc}(\varepsilon_\mathrm{muc}-\bar{\varepsilon}),\\
% \sigma_\mathrm{lig}(\varepsilon_\mathrm{lig})&= \sigma_{\mathrm{lig},0}+E_\mathrm{lig}(\varepsilon_\mathrm{lig}-\bar{\varepsilon}),\\
% \sigma_\mathrm{ta}(\varepsilon_\mathrm{ta},a_\mathrm{ta})&= \sigma_{\mathrm{ta},0}+E_\mathrm{ta}(\varepsilon_\mathrm{ta}-\bar{\varepsilon}),
% \end{split}
\end{equation}
where 
$$\sigma_{j,0}=\bar{\sigma}_{j}(\bar{\varepsilon}),~E_{j}=\mathrm{d}\bar{\sigma}_{j}(\bar{\varepsilon})/\mathrm{d} \varepsilon,~j\in \{\mathrm{muc},\mathrm{lig}\},
$$
$$
\sigma_{\mathrm{ta},0}=\bar{\sigma}_\mathrm{ta}(\bar{\varepsilon},\texttt{a}_\mathrm{ta}),~E_{\mathrm{ta}}=\frac{\mathrm{d}\bar{\sigma}_{\mathrm{ta}}(\bar{\varepsilon},\texttt{a}_\mathrm{ta})}{\mathrm{d} \varepsilon},
$$
and $\bar{\sigma}_{i},~i\in \mathcal{I}$ denotes the nonlinear stress function associated with the $i^\mathrm{th}$ layer.

VF tissues exhibit a highly nonlinear hysteretic viscoelatic behaviour \citep{MinTitzeAlipour95,ChanTitze99}.  The literature is rich in various studies attempting to develop VF constitutive models that capture, at least in part, the complex mechanical behaviors of the VF tissues (see the review study of \citet{Miri14}).  For example, \citep{TitzeAlipour06} proposed a one-dimensional  modified Kelvin model for the VF tissues and laryngeal muscles with nonlinear active and passive stresses to account for tissue viscoelasticity and muscle activation and implemented this constitutive modelling framework in simulations of laryngeal postures \citep{TitzeHunter07}.   \citet{ZhangSiegmundChan06} proposed a constitutive model for the VF cover tissues, which consists of a hyperelastic equilibrium network in parallel with an inelastic,
time-dependent network, and integrated it with an ideal string model to gain insights into the influence of cover tissue mechanical behaviour on phonation fundamental frequency. In a study based on measurements collected from porcine VFs, \citet{MiriHerisTripathyWisemanMongeau13} observed that the collagen fibrils, which are major constituent of the VFs, are rope-shaped, where the geometric characteristics of the fibrils have been incorporated in a hyperelastic mechanical model.  In an attempt to capture the anisotropic properties of the VF lamina propria, \citet{Zhang19}  proposed a structurally-based constitutive model that links the microstructural characteristics of the lamina propria to its macromechanical properties; the proposed model has shown good agreement with biaxial tensile testing measurements. 

Herein, and for simplicity, we assume the constitutive stress-strain relations associated with the nonlinear stresses $\bar{\sigma}_{i}$ to be elastic (functions of strain only) and of  exponential type (see \citet{HunterTitze07}) with symmetry about zero strain. In particular,
$$
\bar{\sigma}_{j}=\mathrm{sign}(\varepsilon)m_{j}(\e^{\lvert n_{j}\varepsilon \rvert}-1),~j\in \{\mathrm{muc},\mathrm{lig}\},
$$
and, in the case of the TA muscle, we include stress induced by muscle activation, resulting in
$$
\bar{\sigma}_{\mathrm{ta}}=\mathrm{sign}(\varepsilon)m_{\mathrm{ta}}(\e^{\lvert n_{\mathrm{ta}}\varepsilon\rvert}-1)+\texttt{a}_\mathrm{ta}\sigma_\mathrm{a,max},
$$
where $m_{i},n_{i},~i\in \mathcal{I},$ are parameters of the constitutive relations, and $\sigma_\mathrm{a,max}$ is the maximum active stress in the TA muscle. Symmetric stress-strain relations are employed herein to account for compressive forces developed in the VF, which have been often dismissed in previous studies of  VF biomechanics. The numerical values of the constitutive relation parameters adopted in this study are listed in Table \ref{tab:LayersProperties}.
\begin{table}
    \centering
       \caption{Numerical values of the geometrical and mechanical properties for each layer in the composite VF model: muc (mucosa), lig (ligament), and ta (thyroarytenoid). Cross-sectional areas are adopted from \citet{TitzeAlipour06}, whereas the parameters $m_{i}$, $n_{i}$, and $\sigma_\mathrm{a,max}$ are tuned to match experimental stress-strain curves from cadaver and canine models presented in  \citet[Figure~2.17,~p.~88]{TitzeAlipour06}.}
    \begin{tabular}{||c| c |c| c||}
    \hline
     \backslashbox{parameter}{layer}  &  muc  & lig & ta \\[0.5ex] 
 \hline\hline
         
    $A_{i}$ [mm\textsuperscript{2}]  &5 &6.1 & 40.9\\
    \hline
      $m_{i}$ [kPa]   &1.5 & 2&1\\
      \hline
      $n_{i}$[-] & 7 &10 &8 \\
      \hline
      $\sigma_\mathrm{a,max}$ [kPa]&-&-& 105\\
      \hline
    \end{tabular}

    \label{tab:LayersProperties}
\end{table}

The normal forces in the VF layers are computed as
$
N_{i}=b\int_{-d_{i}/2}^{d_{i}/2}\sigma_{i} \mathrm{d}y_{i},~i\in \mathcal{I}.
$
Substituting Equations \eqref{eq:StressField} and \eqref{eq:StrainField} in  yields
\begin{equation}\label{eq:NormalForces}
     N_{i}=F_{i,0}+(1+\bar{\varepsilon})E_{i}A_{i}\bar{u}'_{i},~i\in \mathcal{I},
\end{equation}
where
\begin{equation}\label{eq:NominalForces}
F_{i,0}=A_{i}\sigma_{i,0},~i\in \mathcal{I},
\end{equation}
denote the nominal normal forces generated by each layer.
% \begin{align}
% N_\mathrm{muc}&=\sigma_{\mathrm{muc},0}A_\mathrm{muc}+(1+\bar{\varepsilon})E_\mathrm{muc}A_\mathrm{muc}\bar{u}'_\mathrm{muc},\\
% N_\mathrm{lig}&=\sigma_{\mathrm{lig},0}A_\mathrm{lig}+(1+\bar{\varepsilon})E_\mathrm{lig}A_\mathrm{lig}\bar{u}'_\mathrm{lig},\\
% N_\mathrm{ta}&=\sigma_{\mathrm{ta},0}A_\mathrm{ta}+(1+\bar{\varepsilon})E_\mathrm{ta}A_\mathrm{ta}\bar{u}'_\mathrm{ta}.
% \end{align}
The total internal normal force is then
\begin{equation}\label{eq:TotalNormalForce}
N=\sum_{i\in \mathcal{I}}N_{i}.
\end{equation}
 
The moment about the center of the $i^\mathrm{th}$ layer due to the stress developed in that layer is given by
$
 M_{i}=-b\int_{-d_{i}/2}^{d_{i}/2} y_{i}\sigma_{i} \mathrm{d}y_{i},~i\in \mathcal{I}.
$
Substituting Equations \eqref{eq:StressField} and \eqref{eq:StrainField} into this formula gives
\begin{equation}\label{eq:CentralMoments}
    M_{i}=(1+\bar{\varepsilon})E_{i}I_{i}w'',~i\in \mathcal{I},
\end{equation}
%
% \begin{align}
% M_\mathrm{muc}&=(1+\bar{\varepsilon})E_\mathrm{muc}I_\mathrm{muc}w'',\\
% M_\mathrm{lig}&=(1+\bar{\varepsilon})E_\mathrm{lig}I_\mathrm{lig}w'',\\
% M_\mathrm{ta}&=(1+\bar{\varepsilon})E_\mathrm{ta}I_\mathrm{ta}w'',
% \end{align}
where 
$
I_{i}=b\int_{-d_{i}/2}^{-d_{i}/2} y_{i}^2 \mathrm{d}y_{i},~i\in \mathcal{I},
$
denotes the area moment of inertia of the $i^\mathrm{th}$ layer.

Let the positions of the geometric centers of the VF layers along the $r$-axis be denoted $r_{i},~i\in \mathcal{I}$; that is, 
\begin{equation}
\begin{split}
r_\mathrm{muc}&=d_\mathrm{ta}+d_\mathrm{lig}+\frac{d_\mathrm{muc}}{2},\\
r_\mathrm{lig}&=d_\mathrm{ta}+\frac{d_\mathrm{lig}}{2},\\
r_\mathrm{ta}&=\frac{d_\mathrm{ta}}{2},
\end{split}
\end{equation}
see Figure \ref{fig:SchematicBeam}. Take a cross-section at longitudinal position $x$ and consider an arbitrary point on the cross-section located at a vertical position $r= r_{c}$  (see Figure \ref{fig:SchematicBeam}).
% \begin{figure}[http]
%     \centering
%     \includegraphics[width=\linewidth]{ForcesAndMoments.pdf}
%     \caption{Schematic diagram of the composite beam model with normal forces and moments generated by the model layers at  position $x$ .}
%     \label{fig:ForcesAndMoments}
% \end{figure}
The moment at $r_{c}$, denoted $M_{c}$, is given by
\begin{equation}\label{eq:Mc}
\begin{split}
M_{c}=&\sum_{i\in \mathcal{I}}M_{i}+(r_{c}-r_{i})N_{i}\\
=& (1+\bar{\varepsilon})(\sum_{i\in \mathcal{I}}E_iI_{i})w''\\
+& (1+\bar{\varepsilon})\sum_{i\in \mathcal{I}}(r_{c}-r_{i})A_{i}E_{i}\bar{u}'_{i}
\\+& \sum_{i\in \mathcal{I}}(r_{c}-r_{i})F_{i,0}.
\end{split}
\end{equation}

\begin{figure}[htpb]
    \centering
    \includegraphics[width=0.5\linewidth]{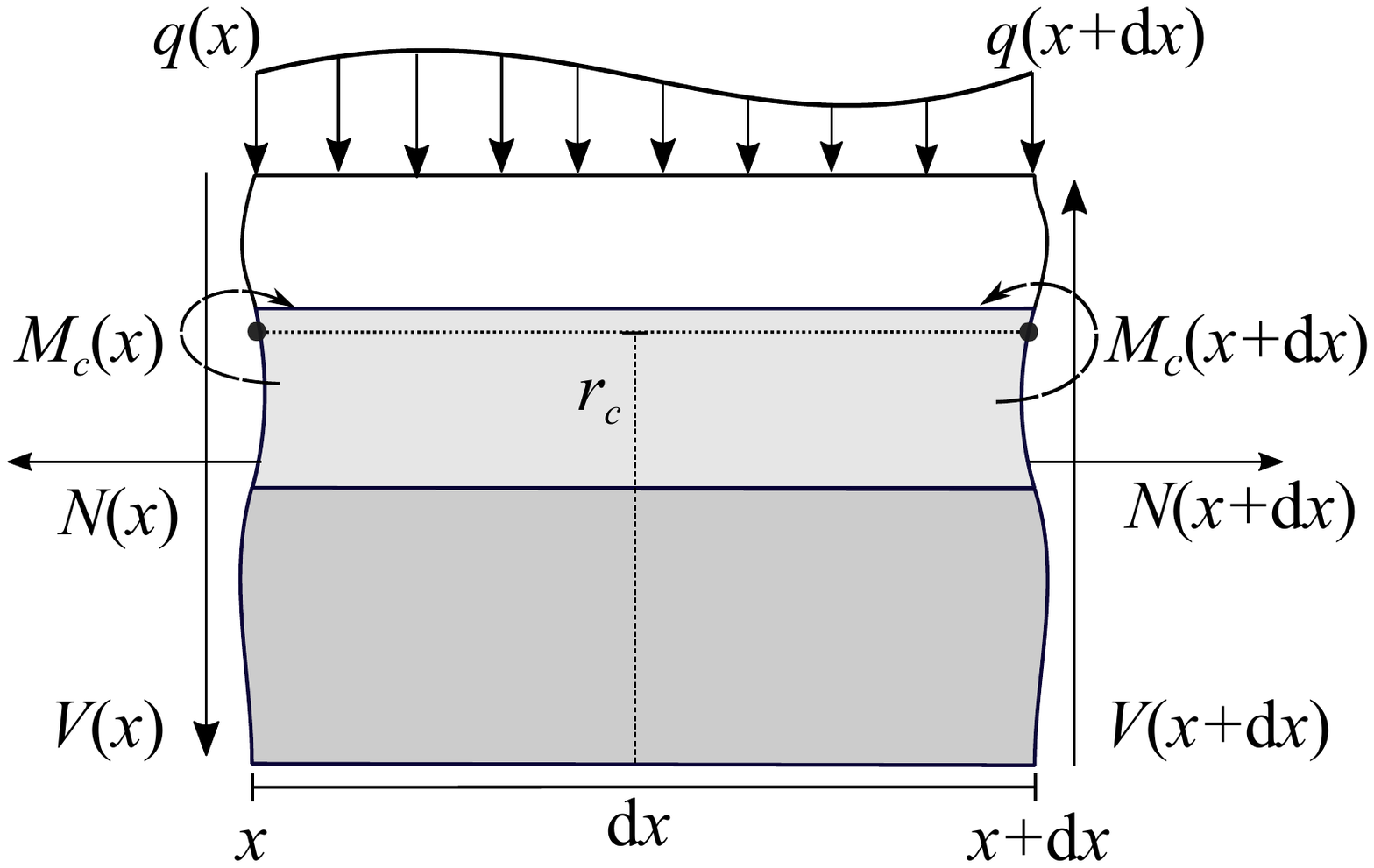}
    \caption{Free-body-diagram of an infinitesimal element of the composite beam VF model.}
    \label{fig:FBDBeam}
\end{figure}

Consider an element of infinitesimal longitudinal length $\mathrm{d}x$ with left edge at position $x$ (see Figure \ref{fig:FBDBeam}), and let $V(x)$ and $q(x)$ denote the shear force and distributed load per unit length, respectively. The  force and moment balances on the infinitesimal element yield
\begin{align}
\label{eq:NormalForceBalance}
N'&=0
, \\\label{eq:ShearForceBalance}
 V'-q&=0,\\\label{eq:MomentBalance}
M'+V&=0,
\end{align}
where second order and higher terms are omitted. 

From Equation \eqref{eq:NormalForceBalance} we deduce that the total normal force $N$ is constant through the VF length. By the assumption that the VF undergoes compression/elongation with associated strain $\bar{\varepsilon}$ (see Equation \eqref{eq:VFElongation}), the force  $N$ should  be equal to the force that results in that strain, which is the sum of nominal forces $F_{i,0},~i\in \mathcal{I},$ (in Equation \eqref{eq:NominalForces}).  That is, 
\begin{equation}\label{eq:ForceAssumption}
N=\sum_{i\in \mathcal{I}}F_{i,0}.
\end{equation}
Therefore, by substituting Equations \eqref{eq:NormalForces} and \eqref{eq:TotalNormalForce} into Equation \eqref{eq:ForceAssumption}, we have
$
\sum_{i\in \mathcal{I}}(1+\bar{\varepsilon})E_{i}A_{i}\bar{u}'_{i}=0,
$
% $$
% (1+\bar{\varepsilon})E_\mathrm{muc}A_\mathrm{muc}\bar{u}'_\mathrm{muc}+(1+\bar{\varepsilon})E_\mathrm{lig}A_\mathrm{lig}\bar{u}'_\mathrm{lig}+(1+\bar{\varepsilon})E_\mathrm{ta}A_\mathrm{ta}\bar{u}'_\mathrm{ta}=0,
% $$
implying
\begin{equation}\label{eq:ZeroForceCondition}
\sum_{i\in \mathcal{I}}E_{i}A_{i}\bar{u}'_{i}=0.
\end{equation}

As we are interested in transverse deflection, we aim to obtain a balance equation solely in terms of  $w$. For convenience, we define 
\begin{equation}
 \begin{split}
    l_\mathrm{lig}&= \frac{1}{2}(d_\mathrm{lig}+d_\mathrm{muc})\\
    l_\mathrm{ta}&= \frac{1}{2}(d_\mathrm{ta}+2d_\mathrm{lig}+d_\mathrm{muc}),\\
l_\mathrm{muc}&=\frac{l_\mathrm{lig}E_\mathrm{lig}A_\mathrm{lig}+l_\mathrm{ta} E_\mathrm{ta}A_\mathrm{ta}}{E_\mathrm{muc}A_\mathrm{muc}+E_\mathrm{lig}A_\mathrm{lig}+E_\mathrm{ta}A_\mathrm{ta}},
\end{split}
\end{equation}
and
\begin{equation}
    \begin{split}
     \alpha_\mathrm{muc}&=-l_\mathrm{muc},\\
     \alpha_\mathrm{lig}&=l_\mathrm{lig}-l_\mathrm{muc},\\
     \alpha_\mathrm{ta}&=l_\mathrm{ta}-l_\mathrm{muc}. 
    \end{split}
\end{equation}
From the continuity condition in Equation  \eqref{eq:ContinuityCondition} and the zero force condition in Equation \eqref{eq:ZeroForceCondition}, it can be deduced that the displacement functions $\bar{u}_{i},~i\in\mathcal{I},$ satisfy the relations 
\begin{equation}\label{eq:UBarToW}
\bar{u}'_{i}=\alpha_{i}w'',~i\in \mathcal{I}.
\end{equation}
Substituting Equation \eqref{eq:UBarToW} into  Equation \eqref{eq:Mc}, we obtain
\begin{equation}\label{eq:McSimplified}
\begin{split}
M_{c}%&\sum_{i\in \mathcal{I}}M_{i}+(r_{0}-r_{i})N_{i}\\
%=& (1+\bar{\varepsilon})(\sum_{i\in \mathcal{I}}E_iI_{i})w''\\
%+& (1+\bar{\varepsilon})\sum_{i\in \mathcal{I}}(r_{0}-r_{i})A_{i}E_{i}\alpha_{i}w''
%\\+& \sum_{i\in \mathcal{I}}(r_{0}-r_{i})A_{i}\sigma_{i,0}\\
&=\mu_{c} w''+M_{c,0},
\end{split}
\end{equation}
where
\begin{equation}\label{eq:Muc}
\mu_{c}=(1+\bar{\varepsilon})\sum_{i\in \mathcal{I}}E_iI_{i}+(r_{c}-r_{i})A_{i}E_{i}\alpha_{i}
\end{equation}
is the composite bending stiffness and
\begin{equation}\label{eq:M0}
M_{c,0}=\sum_{i\in \mathcal{I}}(r_{c}-r_{i})F_{i,0}
\end{equation}
is the nominal moment at $r=r_{c}$ due to the nominal normal forces. Note that the bending stiffness is strain-dependent, which can be of importance in posturing scenarios with large VF strains.

Combining   Equations \eqref{eq:ShearForceBalance} and \eqref{eq:MomentBalance} results in
$$
 M''_{c}+q=0,
$$
which, in terms of the deflection $w$ (obtained by substituting in Equation \eqref{eq:McSimplified}) is 
\begin{equation}\label{eq:BeamEqn}
\mu_{c} w''''+q=0.
\end{equation}
The distributed load $q$ is due to VF contact, which is assumed to be proportional to the transverse overlap beyond the medial plane. That is,
\begin{equation}\label{eq:ContactModel}
q=K_\mathrm{col}(w-x\tan(\theta_{G}))\mathbf{H}(w-x\tan(\theta_{G})),
\end{equation}
where
$K_\mathrm{col}$ is a stiffness coefficient associated with VF contact, $\theta_{G}$ is the clockwise angle between the medial plane and the deformed VF configuration under strain $\bar{\varepsilon}$ (see Figure \ref{fig:SchematicBeam}), and $\mathbf{H}$ is the Heaviside function. 

In this work, we assume zero transverse deflection  at the anterior and posterior ends of the VF. That is,
\begin{equation}\label{eq:ZeroDisplacementBCs}
w(0)=w(L)=0.
\end{equation}
Moreover, we assume zero moment at the posterior VF margin, 
\begin{equation}\label{eq:PosteriorMomentBC}
M_{c}(L)=0. 
\end{equation}
Furthermore, we assume a reactive moment at the anterior VF margin that is proportional to the rotational displacement with respect to the VF angle at rest, $\theta_{0}$. The total angle at the anterior margin between the medial plane and the VF is approximately given by $\theta_{G}-w'(0)$. Consequently, the moment  boundary condition at the anterior margin is given by
\begin{equation}\label{eq:AnteriorMomentBC}
M_{c}(0)=-K_{r}(\theta_{G}-w'(0)-\theta_{0}),
\end{equation}
where $K_{r}$ is a rotational stiffness coefficient. Like the nominal strain $\bar{\varepsilon}$, it is assumed that the angle $\theta_{G}$ is  known a priori. Finally,  we assume that $r_{c}$ corresponds to the geometrical center of the ligament, that is $r_{c}=r_\mathrm{lig}$. This assumption, in addition to the boundary condition given in Equation \eqref{eq:PosteriorMomentBC}, implies that the total normal force $N$ is positioned at the geometrical center of the ligament. This can be deduced from the fact that $M_{c}(L)=(r_{c}-r_\mathrm{N})N$, where $r_\mathrm{N}$ denotes the $r$-position of the total normal force $N$ (that is, the force centroid).

%%%%%%%%%%%%%%%%%%%%%%%%%%%%%%%%%%%%%%%%%%%%%%%%%%%%%%%%%%%%
\section{Analytical insights from a special case}
\label{sec:TheoreticalAnalysis}
To gain simple yet useful insights into how internal moments inside the VF beam model affect its curvature, we consider the scenario of zero contact forces (i.e., $q(x)=0$) and assume $\lvert w'(0) \rvert \ll \lvert\theta_{G} \rvert$, which reduces the boundary condition given in Equation \eqref{eq:AnteriorMomentBC} to
\begin{equation}\label{eq:AnteriorMomentBCSimplified}
  M_{c}(0)= -K_{r}(\theta_{G}-\theta_{0}).
\end{equation}
Equation \eqref{eq:BeamEqn}, with boundary conditions given in Equations \eqref{eq:ZeroDisplacementBCs}, \eqref{eq:PosteriorMomentBC}, and \eqref{eq:AnteriorMomentBCSimplified}, and the definition of $M_{c}$ in Equation \eqref{eq:McSimplified}, can be solved analytically. The curvature of the VF beam model, $w''$, is given explicitly by
\begin{equation}\label{eq:wpp}
w''=-\frac{M_{c,0}}{\mu_{c}}+\left(1-\frac{x}{L}\right)\frac{K_{r}}{\mu_{c}}(\theta_{0}-\theta_{G}).
\end{equation}
Note that positive $w''$ implies a convex VF geometry, whereas negative curvature implies a concave geometry. Recalling the definition of $M_{c,0}$ given in Equation \eqref{eq:M0} and implementing the assumption that $r_{c}=r_\mathrm{lig}$  result in
\begin{align*}
-M_{c,0}&=-\frac{d_\mathrm{ta}+d_\mathrm{lig}}{2}F_{\mathrm{ta},0}+\frac{d_\mathrm{muc}+d_\mathrm{lig}}{2}F_{\mathrm{muc},0}\\
&=\tilde{M}_\mathrm{ta}+\tilde{M}_{\mathrm{muc}},
\end{align*}
where
\begin{equation}\label{eq:TildeMta}
\tilde{M}_\mathrm{ta}=-\frac{d_\mathrm{ta}+d_\mathrm{lig}}{2}F_{\mathrm{ta},0},
\end{equation}
and
\begin{equation}\label{eq:TildeMmuc}
\tilde{M}_{\mathrm{muc}}=\frac{d_\mathrm{muc}+d_\mathrm{lig}}{2}F_{\mathrm{muc},0}.
\end{equation}
By additionally defining 
\begin{equation}\label{eq:TildeMr}
\tilde{M}_{r}={K_{r}}(\theta_{0}-\theta_{G}),
\end{equation}
Equation \eqref{eq:wpp} can be rewritten as
\begin{equation}\label{eq:Wpp2}
w''=\frac{1}{\mu_{c}}\left[\tilde{M}_\mathrm{ta}+\tilde{M}_\mathrm{muc}+\left(1-\frac{x}{L}\right)\tilde{M}_{r}\right].
\end{equation}

In the following discussion, we assume that bending stiffness  $\mu_{c}$ is always positive ($\mu_{c}$ according to Equation \eqref{eq:Muc} changes with the elongation or compression of the VF beam model). First, let us analyze abstractly the effects of the moment terms $\tilde{M}_\mathrm{ta}$, $\tilde{M}_\mathrm{muc}$, and $\tilde{M}_r$ on the VF curvature. We can observe from Equation \eqref{eq:Wpp2} that the effect of the reactive moment $\tilde{M}_{r}$ on the VF curvature decays linearly with a maximum effect (in magnitude) at $x=0$ and zero effect at $x=L$. In contrast, $\tilde{M}_\mathrm{ta}$ and $\tilde{M}_\mathrm{muc}$ have spatially invariant (i.e., constant) effects on $w''$. The curvature is positively correlated with $\tilde{M}_{r}$ and $\tilde{M}_\mathrm{ta}+\tilde{M}_\mathrm{muc}$. That is, $\tilde{M}_{r}>0$ and $\tilde{M}_\mathrm{ta}+\tilde{M}_\mathrm{muc}>0$ implies positive curvature (i.e., convex VF geometry), whereas $\tilde{M}_{r}<0$ and $\tilde{M}_\mathrm{ta}+\tilde{M}_\mathrm{muc}<0$ implies negative curvature (i.e., concave VF geometry). Considering the fact that the effect of the anterior reactive moment $\tilde{M}_{r}$ decays linearly along the VF length and the nominal moments induced by the VF layers, $\tilde{M}_\mathrm{ta}$ and $\tilde{M}_\mathrm{muc}$, are spatially invariant, there can arise an interesting scenario for which the curvature changes sign along the VF length. In particular, when
\begin{equation}\label{eq:HourGlassShapeCondition}
\tilde{M}_{r}>-(\tilde{M}_\mathrm{ta}+\tilde{M}_\mathrm{muc}) > 0,
\end{equation}
$w''$ is positive on $[0,x_\mathrm{cr})$, where
$$
x_\mathrm{cr}=L\left(1+\frac{\tilde{M}_\mathrm{ta}+\tilde{M}_\mathrm{muc}}{\tilde{M}_{r}}\right),
$$   
and negative for $x\in (x_\mathrm{cr},L]$, a change from convexity to concavity. The conditions on the internal moments and resulting VF shapes from this analysis are summarized in Table \ref{tab:Conditions}.  
\begin{table}[htpb]
    \centering
    \caption{Conditions on the moments applied to the VF composite model based on Equation \eqref{eq:Wpp2} and the resulting VF shapes: $\rcurvearrowright$ (convex) and $\curvearrowright$ (concave).}
   \begin{tabular}{||c |c||}
   \hline
     condition   & VF shape \\
         [0.5ex] 
 \hline\hline      $\tilde{M}_\mathrm{muc}+\tilde{M}_\mathrm{ta}$ and $\tilde{M}_{r}> 0$   & 
$\rcurvearrowright
$ \\
   \hline
$\tilde{M}_\mathrm{muc}+\tilde{M}_\mathrm{ta}$ and $\tilde{M}_{r}< 0$ & $
\curvearrowright$\\
   \hline
$\tilde{M}_{r}>-(\tilde{M}_\mathrm{muc}+\tilde{M}_\mathrm{ta})>0$ & $\rcurvearrowright \curvearrowright
$\\\hline
   \end{tabular}
    \label{tab:Conditions}
\end{table}

We note that a convex VF geometry is a defining characteristic of the bowed VF pattern. Moreover, the concave VF geometry can be associated with posterior glottal opening. Furthermore, transition along the VF length from convex to concave resembles the hourglass glottal pattern (see Figure \ref{fig:GlottalShapes}). This demonstrates that the beam model has the capacity to produce the experimentally-observed glottal configurations shown in Figure \ref{fig:GlottalShapes}.

Now, let us relate the findings listed in Table \ref{tab:Conditions} to physiological posturing scenarios. The term $\tilde{M}_{r}$, as seen from its definition in Equation \eqref{eq:TildeMr}, is related to VF adduction and abduction, wherein $\tilde{M}_{r}>0$ corresponds to VF adduction ($\theta_{G}< \theta_{0}$) and $\tilde{M}_{r}<0$ corresponds to VF abduction ($\theta_{G}> \theta_{0}$). The  term $\tilde{M}_\mathrm{muc}+\tilde{M}_\mathrm{ta}$, which is defined according to Equations \eqref{eq:TildeMta} and \eqref{eq:TildeMmuc}, is determined by the reactive moments developed in the VF layers, especially the mucosa and TA muscle, during VF tensioning or compression. %For example, the case when  $\tilde{M}_\mathrm{muc}+\tilde{M}_\mathrm{ta}=0$ corresponds to the scenario when the VF is under zero/minute tension or compression, such that reactive forces in the VF layers are negligible. let us study the cases of substantial  VF tension and compression, 
Note that, based on the area measurements (Table \ref{tab:LayersProperties}) and the assumption of uniform thickness $b$, the moment arm of the TA muscle, $(d_\mathrm{ta}+d_\mathrm{lig})/2$, is larger than that of the mucosa, $(d_\mathrm{muc}+d_\mathrm{lig})/2$. In the case of VF compression,  the compressive forces in the TA muscle are typically larger in magnitude than that in the mucosa (i.e., $F_{\mathrm{ta},0}\ll F_{\mathrm{muc},0}<0$); hence, the moment induced by the TA muscle, $\tilde{M}_\mathrm{ta}$, is positive and predominant making $\tilde{M}_\mathrm{ta}+\tilde{M}_\mathrm{muc}>0$. On the other hand, when the VF is tensioned due to activating the TA muscle, the force  $F_{\mathrm{ta},0}$ is positive and predominant and, consequently, the term $\tilde{M}_\mathrm{ta}$ is negative (see Equation \eqref{eq:TildeMta}) and predominant. This scenario  results in  $\tilde{M}_\mathrm{ta}+\tilde{M}_\mathrm{muc}<0$. These relations between the moment terms and corresponding laryngeal posturing scenarios are summarized in Table \ref{tab:Conditions2}.
\begin{table}[htpb]
\centering
 \caption{Conditions on the moments applied to the VF composite beam model and  corresponding physiological posturing scenarios }
    \begin{tabular}{||c| c||}
    \hline
      condition   & physiological scenario \\
          [0.5ex] 
 \hline\hline      $\tilde{M}_{r}> 0$   & VF adduction 
 \\
    \hline
$\tilde{M}_{r}< 0$ & VF abduction\\
    \hline
$\tilde{M}_\mathrm{muc}+\tilde{M}_\mathrm{ta}>0$ & VF compression\\
\hline
$\tilde{M}_\mathrm{muc}+\tilde{M}_\mathrm{ta}<0$ &  TA muscle activation\\
\hline
    \end{tabular}
    
    \label{tab:Conditions2}
\end{table}

The combined findings presented in Tables \ref{tab:Conditions} and \ref{tab:Conditions2} can be summarized by following observations: The bowed shape with convex VF geometry can be due to
    (a) positive reactive moment at the anterior margin ($\tilde{M}_{r}>0$) during VF adduction, and/or
    (b) internal moments during VF compression, wherein $\tilde{M}_\mathrm{ta}+\tilde{M}_\mathrm{muc}>0$.
Moreover, the concave VF shape arising in the case of posterior glottal opening can be due to 
    (a) negative reactive moment at the anterior margin ($\tilde{M}_{r}<0$) during VF abduction,  and/or
    (b) sufficiently large activation of the TA muscle, wherein $\tilde{M}_\mathrm{ta}+\tilde{M}_\mathrm{muc}<0$. 
The hourglass shape may necessitate a coordinated laryngeal maneuver that involves sufficient TA activation ($\tilde{M}_\mathrm{ta}+\tilde{M}_\mathrm{muc}<0$) and VF adduction ($\tilde{M}_{r}>0$) such that $\tilde{M}_{r}+\tilde{M}_\mathrm{ta}+\tilde{M}_\mathrm{muc}>0$.  
% Before concluding, we need to highlight an interesting observation from Equation \eqref{eq:Wpp2}. By recalling the definition of the bending stiffness term $\mu_{c}$ given in Equation \eqref{eq:Muc}, we observe that $w''$ is inversely proportional to $1+\bar{\varepsilon}$, which corresponds to the VF stretch. This strain dependence of the bending stiffness indicates that when the VF experiences significant elongation ($1+\bar{\varepsilon}\gg 1$), the second derivative of $w$ tends to be small, implying (almost) straight VF layout. On the other-hand, when the VF goes under significant compression ($1+\bar{\varepsilon}$ is close to zero), the second derivative becomes large implying curved VF shape. This suggests that VF geometry tends to be curved under VF compression, and tends to be straight under VF tension. 

% In Section \ref{sec:Simulations}, we further investigate curved VF geometry and incomplete glottal closure by additionally incorporating  the posture model of \citet{TitzeHunter07}.

%%%%%%%%%%%%%%%%%%%%%%%%%%%%%%%%%%%%%%%%%%%%%%%%%%%%%%%%
\section{Simulations of the combined beam and posturing model}
\label{sec:Simulations}
In this section we further investigate VF curvature and incomplete glottal closure by combining our beam model with the VF posturing modeling introduced by \citet{TitzeHunter07}. In particular, we adopt the implementation of \citet{AlzamendiPetersonErathHillmanZanartu21}. The posture model relates activation of the five intrinsic muscles to the prephonatory glottal configuration, and in particular, the rotational and linear displacements of the cricothyroid joints and the arytenoid cartilages, where the VFs and intrinsic muscles are modelled as spring-like elements. From the aforementioned displacements, the VF nominal strain $\bar{\varepsilon}$ and glottal angle $\theta_{G}$ are estimated. Similar to the muscle activation parameter $\mathrm{a}_\mathrm{ta}$ embedded in the VF beam model, the posture model relies on five normalized muscle activation parameters, $a_\mathrm{ta}$, $a_\mathrm{ct}$, $a_\mathrm{lca}$, $a_\mathrm{ia}$, and $a_\mathrm{pca}$, which correspond to the TA, cricothyroid (CT), lateral cricoarytenoid (LCA), interarytenoid (IA), and PCA muscles, respectively. In this study, we assume that the muscle activation parameter $\mathrm{a}_\mathrm{ta}$ embedded in the VF beam model is identical in value to the muscle activation parameter $a_\mathrm{ta}$ in the posturing model ($\mathrm{a}_\mathrm{ta}=a_\mathrm{ta}$).

It is important to  mention that the constitutive relations embedded in the VF beam model are different from those in the posture model implementation adopted from \citet{AlzamendiPetersonErathHillmanZanartu21}. The focus of the current study is to replicate the VF static configurations, wherein we employ experimental stress-strain data based on human and canine samples \citep[Figure~2.17,~p.~88]{TitzeAlipour06} to prescribe the mechanical behaviors of the tissues. The posture model of \citet{AlzamendiPetersonErathHillmanZanartu21} instead focuses on  replicating physiologically accurate posturing and phonation outputs; this required ad hoc tuning of some posturing model parameters. Prior experimental and numerical studies typically suffer from significant variability in the reported numerical values of biomechanical parameters (see, e.g., \citet{TitzeAlipour06,TitzeHunter07,HunterTitze07,PalaparthiSmithTitze19}), and in some cases numerical values are missing altogether, which  motivates the ad hoc tuning approach adopted by \citet{AlzamendiManriquezHadwinDengPetersonErathMehtaHillmanZanartu20}.
 
The posturing model in \citet{TitzeHunter07,AlzamendiPetersonErathHillmanZanartu21} is dynamic due to inertial and viscous effects. In this study, and as we are interested in static posturing scenarios, the posture model is run until the VF strain and glottal angle reach steady-state and these values are input into the composite beam model. Once the VF strain and glottal angle parameters are fed into the beam model, Equation \eqref{eq:BeamEqn}, supplemented with Equations \eqref{eq:Mc}, \eqref{eq:ContactModel}, \eqref{eq:ZeroDisplacementBCs}, \eqref{eq:PosteriorMomentBC}, and \eqref{eq:AnteriorMomentBC}, is solved numerically. The aforementioned equations and boundary conditions are discretized by means of finite difference. For the simulations $\theta_{0}$ is set as the glottal angle from the posturing simulations when all laryngeal muscles are inactive. Numerical values for the remaining VF beam model parameters are listed in Table \ref{tab:SomeParametersBeam}.
\begin{table}
    \centering
    \caption{Numerical values of some of the VF beam model parameters.}
    \begin{tabular}{||c|c||}
\hline       parameter  & numerical value \\
         [0.5ex] 
 \hline\hline $L_{0}$ [mm] & 15 \\
       \hline $b$ [mm] & 5 \\
    \hline    $\theta_{0}$ [rad] & 0.2540 \\
    \hline    $K_\mathrm{col}$ [N/m\textsuperscript{2}] & $2\times 10^{8}$ \\
\hline        $K_{r}$ [N.m] & 0.05 \\
\hline
    \end{tabular}
    \label{tab:SomeParametersBeam}
\end{table}

To clearly illustrate the glottal geometries resulting from the simulations, a coordinate system $(x_{1},x_{2})$ with origin at the anterior margin of the VFs is established. The $x_{1}$-axis is aligned along the medial plane pointing in the posterior direction and the $x_{2}$-axis is perpendicular to the medial plane pointing to the right, relative to the human body frame (see Figure \ref{fig:SchematicBeam}). In all figures presented in this section the VF configurations are plotted with respect to this coordinate system; model symmetry is utilized to produce the opposing VF shape.     

This section explores several laryngeal maneuvers and how they influence the VF geometry\footnote{The end points (anterior and posterior margins) of the VFs resulting from the proposed beam model are identical to those established by the posturing model of \citet{TitzeHunter07}. Accounting for internal bending moments results in deviation of the VF shape from the linear medial surface prescription (with angle $\theta_{G}$) of \citet{TitzeHunter07}.}. In particular, and motivated by previous clinical, experimental, and numerical findings \citep{YinZhang14,MorrisonRammage93, ChhetriNeubauer15,YinZhang16}, we consider laryngeal maneuvers associated with adductory (TA, LCA, and IA) and abductory (PCA), muscles as they have been found to play major roles in inducing curved glottal geometries. The CT muscle has been found to play a major role in regulating phonation fundamental frequency by stretching the VFs, but not in posturing and is thus excluded from this study.   Herein, we compare simulation results with findings from previous clinical, experimental,  and high-fidelity numerical studies to verify the proposed VF beam model. Moreover, we attempt to elucidate potential mechanisms underlying the curved VF geometries observed clinically by analyzing the beam model details (see Figure \ref{fig:GlottalShapes}).

\begin{figure*}[http]
    \centering
    \includegraphics[width=0.9\linewidth,trim={0.8in 0in 0.8in 0in}]{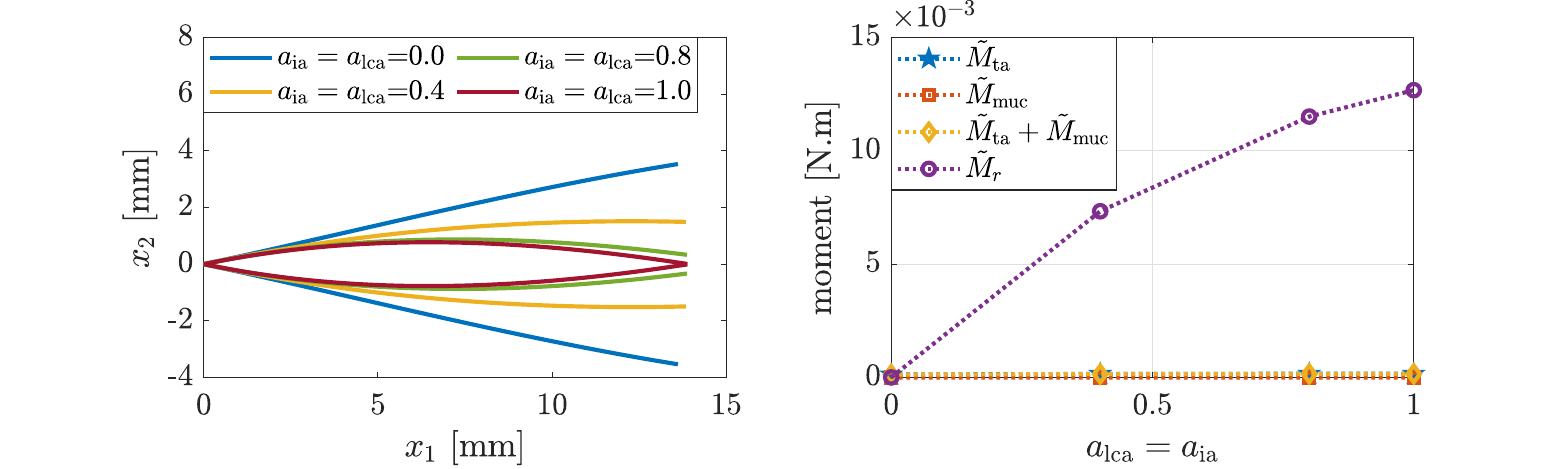}
    \caption{ (Left) glottal profile and (right) induced moments  for increasing LCA and IA activation levels and other intrinsic muscles being inactive.}
    \label{fig:Vary_LCA_IA}
\end{figure*}

% \begin{enumerate*}[label={(\arabic*)}]
%     \item ,
%     \item ,
%     \item ,
%     \item , and
%     \item increasing  activation of the PCA muscle, while the CT muscle is inactive and the adductors are kept at constant muscle activation levels corresponding to near full adduction. 
%   \end{enumerate*}  
First, we investigate the effects of increasing co-activation of the LCA and IA muscles, which are responsible for adducting the VFs \citep{AlzamendiPetersonErathHillmanZanartu21}, while the remaining intrinsic muscles are inactive. Figure \ref{fig:Vary_LCA_IA} presents the glottal shapes and the induced moments corresponding to simulations wherein LCA and IA muscle activation levels are increased simultaneously. Figure \ref{fig:Vary_LCA_IA}(left) shows that co-activation of the LCA and IA muscles leads to posterior glottal closure with a remaining   mid-membranous gap. This convex   VF shape matches previous clinical and numerical findings  \citep{ChhetriNeubauer15,YinZhang16}. Figure \ref{fig:Vary_LCA_IA}(right) shows that the VF convexity is due to the predominance of the reactive moments at the anterior VF margin ($\tilde{M}_{r}$ is positive and relatively large), which arises due to VF adduction $(\theta_{G}<\theta_{0})$, which agrees with the theoretical predictions in Section \ref{sec:TheoreticalAnalysis}.

\begin{figure*}[http]
    \centering
    \includegraphics[width=0.9\linewidth, trim={0.8in 0in 0.8in 0in}]{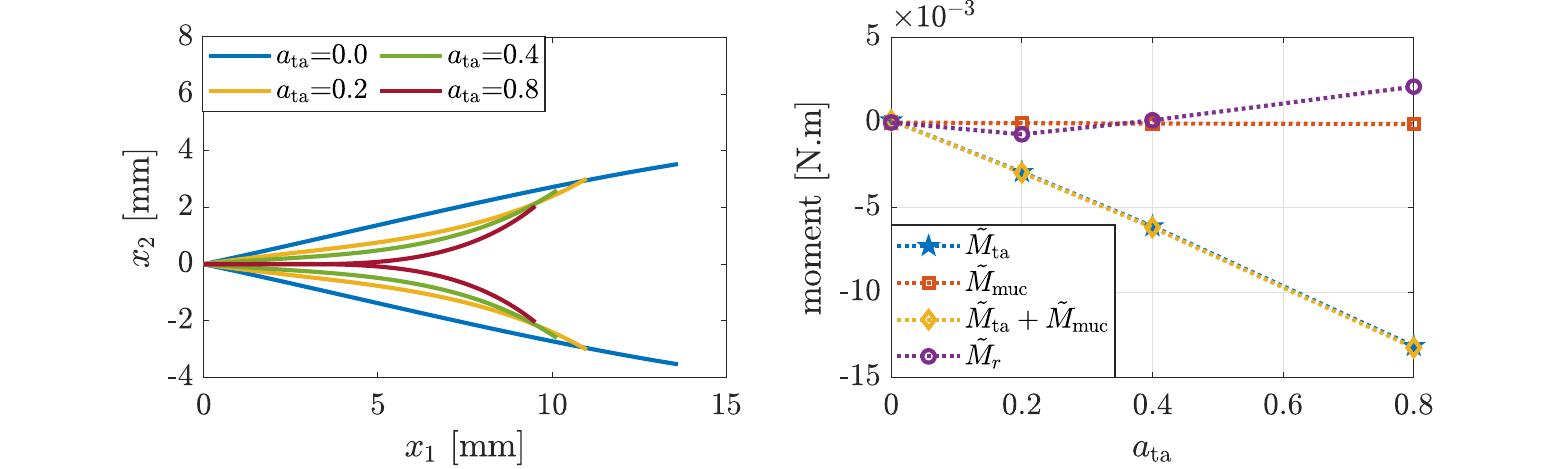}
    \caption{(Left) glottal profile and (right) induced moments  for increasing TA activation levels and other intrinsic muscles being
inactive.}
    \label{fig:Vary_TA}
\end{figure*}

Figure \ref{fig:Vary_TA} exhibits the glottal shapes and induced moments corresponding to simulations wherein TA muscle activation levels are increased, while all other intrinsic muscles  are inactive. Figure \ref{fig:Vary_TA}(left) shows that isolated activation of the TA muscle leads to anterior and mid-membranous glottal closure with remaining  posterior opening, while also shortening the folds. The resulting concave VF shapes are in agreement with previous experimental and numerical investigations \citep{ChhetriNeubauer15,YinZhang16}. Figure \ref{fig:Vary_TA}(right) shows that the concavity is primarily determined by the internal moments induced by the TA muscle activation ($\tilde{M}_{\mathrm{ta}}$ is negative and relatively large in magnitude), which is in alignment with the analysis in Section \ref{sec:TheoreticalAnalysis}.  

\begin{figure*}[http]
    \centering
    \includegraphics[width=0.9\linewidth, trim={0.8in 0in 0.8in 0in}]{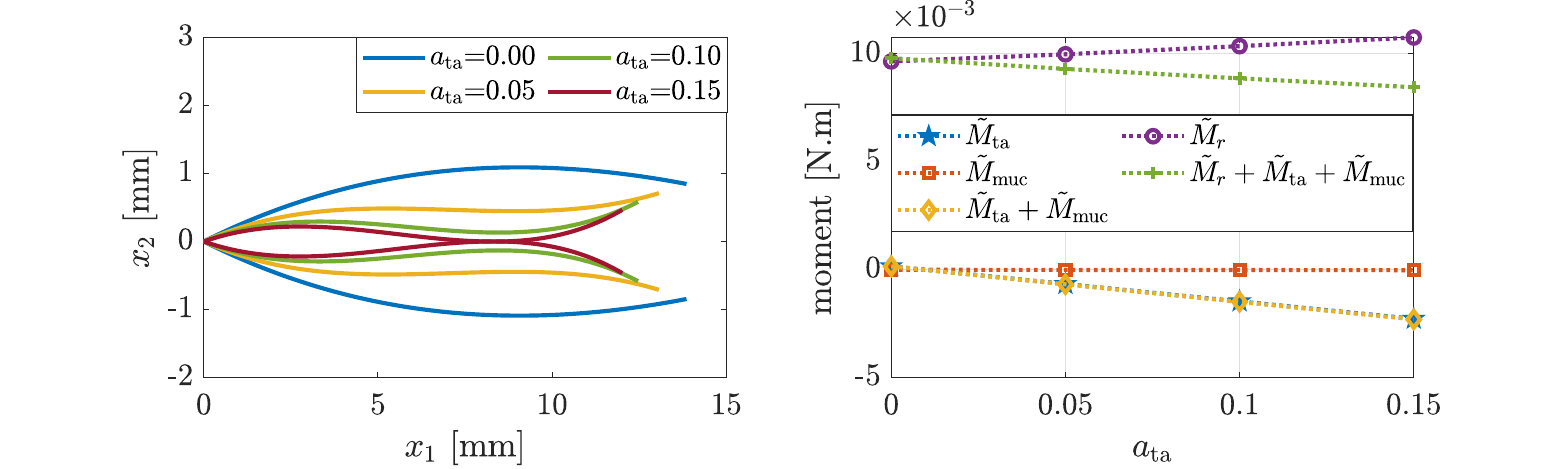}
    \caption{ (Left) glottal shapes, and (right) induced moments  for increasing TA activation levels and other intrinsic muscle activation levels being at $a_\mathrm{lca}=a_\mathrm{ta}=0.6$ and $a_\mathrm{ct}=a_\mathrm{pca}=0$.}
    \label{fig:Vary_TA_Hour_Glass}
\end{figure*}

In an effort to explore the mechanics of the hourglass glottal shape, we explore the glottal shape associated with increasing activation of the TA muscle while the LCA and IA are kept at constant non-zero activation levels. Simulating such maneuvers is encouraged by the findings from the theoretical analysis in Section \ref{sec:TheoreticalAnalysis} and the experimental observations in \citet{ChhetriNeubauerBerry12}. Figure \ref{fig:Vary_TA_Hour_Glass} displays the glottal shapes and induced moments associated with slight increasing activation of the TA muscle, while the LCA and IA are kept at constant  levels ($a_\mathrm{lca}=a_\mathrm{ta}=0.6$). Figure \ref{fig:Vary_TA_Hour_Glass}(left) shows  that in the case of zero TA activation, the glottal shape is bowed with slight, but not full, posterior adduction. As TA activation is increased, a medial bulge is observed whereas anteriorly the glottal geometry is still convex, resulting in an overall hourglass shape. Figure  \ref{fig:Vary_TA_Hour_Glass}(right) displays how the internal moments $\tilde{M}_{\mathrm{ta}}$ and  $\tilde{M}_{\mathrm{muc}}$ and the reactive moment at the anterior margin $\tilde{M}_{r}$ satisfy the  condition  of Equation \eqref{eq:HourGlassShapeCondition}. This aligns with the analysis in Section \ref{sec:TheoreticalAnalysis} and suggests that the hourglass glottal shape necessitates involvement of reactive moments at the anterior VF margin (associated with VF adduction) and  internal moments induced inside the VF layers (primarily the TA muscle). In addition, this finding is in good agreement with observations in \citet{ChhetriNeubauerBerry12}, which showed that an hourglass shape is induced by coactivating  all the adductory muscles.

\begin{figure}[htpb]
    \centering
    \includegraphics[width=0.45\linewidth]{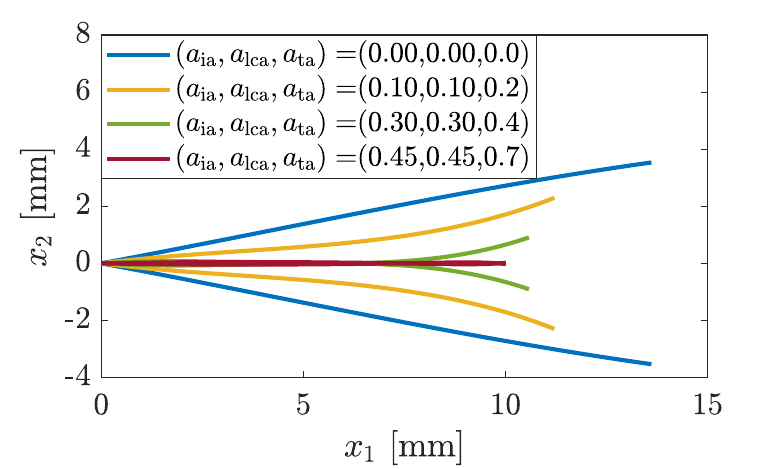}
    \caption{Glottal shapes for increasing co-activation levels of all adductory muscles, and the PCA and CT muscles being inactive.}
    \label{fig:Vary_Add}
\end{figure}

 In aggregate, Figures~\ref{fig:Vary_LCA_IA}-\ref{fig:Vary_TA_Hour_Glass} indicate that dismissing the TA muscle or the LCA and IA muscles cannot produce full glottal closure, therefore, in the next set of simulations, we investigate the effects of co-activating all of the adductory muscles. Figure \ref{fig:Vary_Add} shows that (almost) full glottal closure can be attained when all adductors are co-activated ($a_\mathrm{ia}=a_\mathrm{lca}=0.45,~a_\mathrm{ta}=0.7$), which is in alignment with previous  experimental and numerical investigations \citep{ChhetriNeubauer15,YinZhang16}.

% In the following set of simulations, we  consider the influence of increasing activation of the CT muscle, which is a primary regulator of phonation fundamental frequency \citep{ChhetriNeubauerSoferBerry14}, on the glottal configuration.  Figure \ref{fig:Vary_CT} displays the glottal patterns associated with increasing activation of the CT muscle, where all other laryngeal muscles are kept inactive. It is observed that as CT activation increases, the VFs get slightly adducted with concave geometry. Unfortunately, these results are not consistent with clinical findings as increasing CT activation typically results in abducted VFs, with convex geometry \citep{ChhetriNeubauer15,YinZhang16}.  This highlights the limitations of the proposed composite beam model and associated boundary conditions, which we further discuss in the next section.
% \begin{figure}[h]
%     \centering
%     \includegraphics[width=\linewidth]{Vary_CT.eps}
%     \caption{Glottal shapes for increasing activation levels of the CT muscle and the other muscles  muscles being inactive.}
%     \label{fig:Vary_CT}
% \end{figure}

\begin{figure}[htpb]
    \centering
    \includegraphics[width=0.45\linewidth]{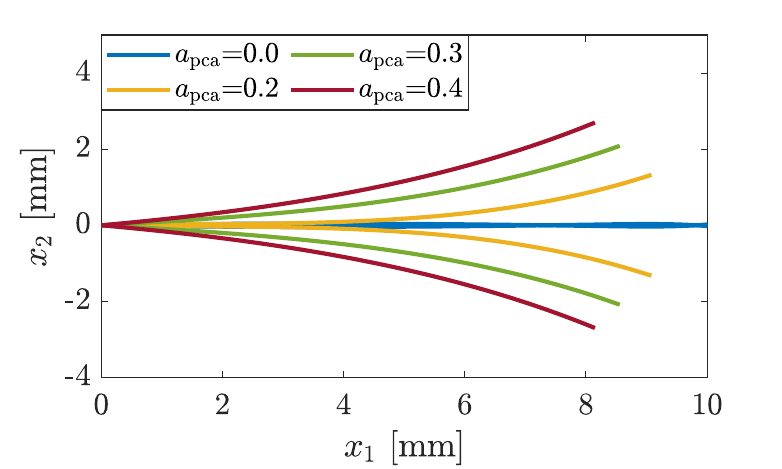}
    \caption{Glottal shapes for increasing activation of the PCA muscle with the activation levels of the other muscles being $(a_\mathrm{lca},a_\mathrm{ia},a_\mathrm{ta})=(0.45,0.45,0.7)$ and $a_\mathrm{ct}=0$.}
    \label{fig:Vary_PCA}
\end{figure}

Finally, we explore the effects of the PCA muscle, a primary VF abductor, on the glottal geometry, where we consider simulations motivated by the clinical observations highlighted in \citet{MorrisonRammage93}. Figure \ref{fig:Vary_PCA} displays glottal patterns associated with increasing PCA activation where adductory muscles are kept at activation levels associated with near full closure. The figure displays that increasing PCA activation leads to posterior opening, while the VFs are sustaining concave shapes similar to those presented when TA alone is activated (see Figure \ref{fig:Vary_TA}). This suggests that PCA activation tends to neutralize the posterior adductory effects of the LCA and IA muscles, which supports the clinical observations highlighted in \citet{MorrisonRammage93}.

%%%%%%%%%%%%%%%%%%%%%%%%%%%%%%%%%%%%%%%%%%%%%%%%%%%%%%%%%%%%%%%%%%%%%%%%%%%%%%%%%%
\section{Discussion}
\label{sec:Discussion}
The results of Sections \ref{sec:TheoreticalAnalysis} and \ref{sec:Simulations} highlight potential mechanisms underlying different patterns of incomplete glottal closure. In particular, results indicate that bowed VF shapes result, in part, from low or null activation of the TA muscle in combination with co-activation of the LCA and IA muscles.
%, agreeing with previous  clinical and numerical findings \citep{ChhetriNeubauer15,YinZhang16}.
The predominant mechanism in this case is the anterior reactive moments that resists bringing the VFs together during adduction. This pattern can also arise in the case of low TA muscle activation and VF compression, as suggested by the analysis in Section \ref{sec:TheoreticalAnalysis}. In this case, the internal moment induced by the TA muscle tissue ($\tilde{M}_\mathrm{ta}$ is positive and predominantly large) is the driving factor. This scenario (bowing due to VF compression) can potentially take place when extrinsic laryngeal muscles are excessively activated, especially those associated with VF compressing, such as the thyrohyoid muscle \citep{HongYeKimKevorkianBerke97}.   

In addition, our analysis suggests that posterior glottal opening with combined VF concavity results from high activation of the TA muscle and low or null activation of the LCA and IA muscles. Our model suggests that the driving mechanism here is the internal moment induced by the TA muscle activation $\tilde{M}_\mathrm{ta}$, which is negative and predominantly large in magnitude in this case. A similar glottal pattern also occurs when all adductory muscles are activated in addition to the activation of the PCA muscle. This supports the hypothesis of \citet{MorrisonRammage93}, regarding the excessive activation of the PCA muscle in patients with MTD. Finally, our analysis suggests that the hourglass glottal shape may emerge from laryngeal maneuvers that involve, for example, moderate co-activation of all adductory muscles, where both anterior reactive moment and internal moment due to TA muscle activation are at play and opposing each other.  

The implications above concerning potential connections between incomplete and curved glottal closure patterns and particular muscular executions may help  speech therapists to uncover the underlying laryngeal mechanisms associated with some voice disorders.  As highlighted in the introduction, incomplete glottal closure can be linked to voice disorders that are characterized by excessive, imbalanced, or deficient activity of the intrinsic and extrinsic muscles such as MTD and Parkinson's disease.  Our analysis in the current work suggests two potential mechanisms underlying  bowed VFs in  some patients with voice disorders (1) the TA muscle is not properly activated (possibly due to muscle activation imbalance), and (2) excessive activation of extrinsic neck muscles, leading to VF compression. Besides, our analysis posits  that in patients with abnormal posterior glottal opening and concave VF geometry either (1)  insufficiently activate the  LCA and IA muscles, whereas the TA muscle is activated sufficiently (in comparison to normal posturing scenarios), or (2) suffer from  excessive activation of all adductory and abductory muscles, where the PCA muscle activation mitigates the effects of the LCA and IA muscles, in agreement with the postulation in \citet{MorrisonRammage93} concerning patients with MTD.  In summary, speech clinicians and therapists may consider the aforementioned candidate underlying mechanisms  of curved and incomplete glottal closure patterns when examining patients with voice disorders involving muscular inefficiencies/deficiencies.
 
A number of simplifying assumptions are embedded in the presented model of this study, including (1) negligible shear deformation,  (2) negligible elastic forces from the  connective tissues attached to the TA muscle, (3) negligible motions in the superior-inferior direction, (4) neglecting potential bending effects from the vocal ligament by setting $r_{c}=r_\mathrm{lig}$, (5) zero moments at the posterior ends of the VFs, (6) small transverse VF deflections, and (7) one-way coupling between the VF beam model and the posturing model, where any contact forces emerging due to the VF curvature do not alter the mechanics of the laryngeal cartilages.

Assumption (1) is a consequence of the adopted Euler-Bernoulli framework. Note that with the uniform thickness assumption, and considering the model dimensions given in Tables \ref{tab:LayersProperties} and \ref{tab:SomeParametersBeam}, the total VF depth, $\sum_{i\in \mathcal{I}}d_{i}$, is approximately 10 mm whereas the resting VF length is 15 mm; hence, the depth and length dimensions are quite comparable. For such cases (thick beams), Timoshenko beam theory \citep{Timoshenko21} is typically adopted to account for shear stresses\footnote{ It is worth noting that the two theories (Euler-Bernoulli and Timoshenko) do coincide for a  uniform homogeneous simply-supported linear beam with specified moments at the end points and zero distributive load, regardless of the beam thickness or mechanical properties (in Section  \ref{sec:TheoreticalAnalysis}, we studied a similar simply-supported case with zero distributive load). These two theories, when compared, tend to produce qualitatively, but not necessarily quantitatively, similar predictions (see, e.g., \citet{BeckDaSilva11}).}. As the goal of the current work is to construct a simple analytically-tractable model that predicts qualitatively the curved glottal configurations observed clinically, we adhered to the Euler-Bernoulli framework, leaving derivations of  more complex models to future work.   We posit that assumption (2) is reasonable as the elastic forces from the connective tissues are passive, mostly only restricting the extent to which the VF deflects. Moreover, assumption (3) is suitable as the majority of the VF motion during posturing occurs medially and/or laterally (see, e.g., the findings of \citet{ChhetriNeubauer15}). Regarding assumption (4), the ligament is stiffer than other VF layers and it geometrically forms the intermediate VF layer, making it the `chassis' of  the VF layered structure; hence setting $r_{c}=r_\mathrm{lig}$, which  indicates that the total normal force in the VF  is positioned at the center of the ligament (see the end of Section \ref{sec:ModelDerivation}), is sensible.  Assumptions (5)-(7), in addition to other assumptions such as the rectangular geometries of the VF layers, are introduced to primarily simplify our analysis; hence, further investigation is needed to verify the validity of such assumptions in different posturing scenarios, and refine them when needed.   
   
Despite these simplifying assumptions, our modelling framework is capable of predicting some of the glottal patterns observed in previous clinical and high-fidelity numerical studies, which is encouraging. Still, the speculations and potential explanations provided in this work need further extensive investigation into the biomechanics of  VF posturing in both healthy subjects and patients with imbalances or deficiencies in the laryngeal muscles. 

%%%%%%%%%%%%%%%%%%%%%%%%%%%%%%%%%%%%%%%%%%%%%%%%%%%%%%%%%%%%%%%%%%%%%%%%%%%%%%%%%%
\section{Conclusion}
\label{sec:Conclusion}
In this study, we introduced a simple one-dimensional Euler-Bernouilli-type composite beam model of the vocal folds to understand the mechanisms underlying glottal configuration and incomplete glottal closure. The model, despite its simplicity, was capable of predicting several clinically observed glottal configurations. Our analysis highlighted how the different patterns of incomplete glottal closure can arise naturally due to the layered VF structure and the associated induced moments.  We coupled the proposed beam model  with the posturing model of \citet{TitzeHunter07} to gain physiologically relevant insights into the role of laryngeal muscle activation. Our analysis showed that a bowed VF shape can arise due to activation of the LCA and IA muscles without incorporating the TA muscle during adduction or due to VF compression. On the other hand, isolated activation of the TA muscle results in medial bulging and posterior glottal opening. Posterior opening can also occur due to activating all adductors in addition to activating the PCA muscle. Moreover, our analysis suggested that an hourglass glottal shape can arise from specific laryngeal maneuvers involving the adductory laryngeal muscles. These results provided potential explanations and conjectures regarding the posturing mechanics of patients with voice disorders such as MTD. 

In future efforts we aim to refine our modelling framework, where two-way coupling between the VF beam model presented herein and the posturing model of \citet{TitzeHunter07} is incorporated, to account for potential effects that curved VF geometries may exert on the mechanics of  laryngeal cartilages. Moreover, we intend to incorporate the beam model with numerical phonation models, to study how curved and partially closed glottal geometries may influence tissue-flow-acoustic interactions, voice quality, and vocal function during phonation.

%%%%%%%%%%%%%%%%%%%%%%%%%%%%%%%%%%%%%%%%%%%%%%%%%%%%%%%%%%%%%%%%%%%%%%%
 \section*{Acknowledgments}
Research reported in this work was supported by the NIDCD of the NIH under Award No. P50DC015446, and ANID BASAL FB0008. The content is solely the responsibility of the authors and does not necessarily represent the official views of the National Institutes of Health.

%%%%%%%%%%%%%%%%%%%%%%%%%%%%%%%%%%%%%%%%%%%%%%%%%%%%%%%%%%%%%%%%%%%%%%%%%%%%%
\section*{Declaration of competing interest}
Mat\'ias Za\~nartu has a financial interest in Lanek SPA, a company focused on developing and commercializing biomedical devices and technologies. His interests were reviewed and are managed by the Universidad T{\'e}cnica Federico Santa Mar\'ia in accordance with its conflict-of-interest policies.

%%%%%%%%%%%%%%%%%%%%%%%%%%%%%%%%%%%%%%%%%%%%%%%%%%%%%%%%%%%%%%%%%%%%%%%%%%%%%%%%
\section*{Author contributions} 
Mohamed Serry: Conceptualization, Methodology, Formal analysis, Software, Visualization, Writing - original draft preparation. Gabriel Alzamendi: Software, Validation, Writing - reviewing and editing. Mat\'ias Za\~nartu: Validation, Writing - reviewing and editing, funding acquisition. Sean Peterson: Supervision, Validation, Writing - reviewing and editing, funding acquisition.

 \bibliographystyle{apalike}
%\bibliography{References}

\end{document}